\documentclass[12pt]{article}
\usepackage{amssymb}
\usepackage{graphicx}
\usepackage[pdfborder={0 0 0},colorlinks=true,allcolors=blue]{hyperref}

\newcommand\pubnumber{Cavendish-HEP-18/01}
\newcommand\pubdate{\today}

\def\institute{Cavendish Laboratory, University of Cambridge, \\ CB3 0HE, Cambridge, UK}
\def\support{\footnote{Work supported by the UK STFC grants ST/L002760/1 and ST/K004883/1 and by the European Research Council Consolidator Grant ``NNLOforLHC2".}}

\def\Title#1{\begin{center} {\Large #1 } \end{center}}
\def\Author#1{\begin{center}{ \sc #1} \end{center}}
\def\Address#1{\begin{center}{ \it #1} \end{center}}

\newcommand\pubblock{\rightline{\begin{tabular}{l} \pubnumber\\
         \pubdate  \end{tabular}}}
\newenvironment{Abstract}{\begin{quotation}  }{\end{quotation}}
\newenvironment{Presented}{\begin{quotation} \begin{center} 
             PRESENTED AT\end{center}\bigskip 
      \begin{center}\begin{large}}{\end{large}\end{center} \end{quotation}}
\def\Acknowledgements{\bigskip  \bigskip \begin{center} \begin{large}
             \bf ACKNOWLEDGEMENTS \end{large}\end{center}}

\begin{document}
\begin{titlepage}
\pubblock

\vfill
\Title{Top Quark Pair Production: theory overview}
\vfill
\Author{Andrew S.~Papanastasiou\support}
\Address{\institute}
\vfill
\begin{Abstract}
In this talk we present the current status of the theoretical description of the 
production of top-quark pairs at the LHC, detailing recent progress for predictions
at the stable-top level as well as for predictions that include the decay of the 
top quarks. 
\end{Abstract}
\vfill
\begin{Presented}
$10^{th}$ International Workshop on Top Quark Physics\\
Braga, Portugal,  September 17--22, 2017
\end{Presented}
\vfill
\end{titlepage}
\def\thefootnote{\fnsymbol{footnote}}
\setcounter{footnote}{0}

\section{Stable tops at high precision} \label{sec:stable}

\vspace{-0.2cm}
With the calculation of higher-order QCD and EW corrections, in addition to
the inclusion of various resummations, the production of a pair of onshell 
stable top-quarks is one of the LHC processes known with highest theoretical precision.  

Predictions probing event kinematics in the multi-TeV regime 
for the $p_{T,t}$, $M_{t\bar{t}}$, $y_t$ and $y_{t\bar{t}}$ distributions 
are all known through NNLO in QCD \cite{Czakon:2015owf,Czakon:2016ckf,Czakon:2016dgf}. 
An important study that these calculations have enabled is the choice of 
dynamical factorization and renormalization scales, which becomes particularly
crucial to describe these regions.
The choice of scale is certainly not unique, may be different for each 
distribution studied, and the best choice is often the subject of lively debate. 
However, when deciding which of two scales is more suited for a given distribution
in fixed-order perturbation theory, using the guiding principle of 
`fastest perturbative convergence', namely picking the scale that leads to 
smallest $K$-factors, as done in ref.~\cite{Czakon:2016dgf} seems perfectly 
reasonable. 
Reassuringly, the authors also find that of the set scales investigated, the scale resulting
in fastest perturbative convergence also leads to the smallest scale uncertainty bands. 
The findings of this study are that for the four principle distributions
known through NNLO,  the optimal choices of central scales are 
$\mu=H_T/4$ for $M_{t\bar{t}}$, $y_t$ and $y_{t\bar{t}}$ and 
$\mu=m_T/2$ for $P_{T,t}$.  
These choices now form the default choices for the NNLO predictions of 
these distributions at the LHC. 
When examining more extreme regions of phase space, or more complex observables, 
such as double-differential cross sections, the study of scale choices  
may require revisiting. 

Despite the impressive nature of the NNLO calculations, typical run-times of $\mathcal{O}(10^5)$ 
CPU-hours per setup mean that the full physics potential of these predictions
may be difficult to reach. 
This problem becomes particularly acute when using these predictions
for PDF-fitting, where often tens of thousands of re-evaluations of the partonic 
cross section with different PDFs are required.  
A major development on this front has been the interfacing of \texttt{Stripper} \cite{Czakon:2014oma} 
to \texttt{fastNLO} \cite{Britzger:2012bs}. 
This allows for tables to be produced for the $t\bar{t}$ process, for any distribution
with fixed bins and a fixed functional form for the scale \cite{Czakon:2017dip}. 
Using these, obtaining the NNLO prediction for a given distribution for any desired PDF choice
is reduced to $\mathcal{O}({\rm seconds})$.

Further progress has been made at the stable-top level by combining NNLO-QCD with 
NLO-EW corrections \cite{Czakon:2017wor,Czakon:2017lgo}. 
These were discussed in a dedicated talk \cite{pagani-top}. 
We highlight here that 
particularly stark EW-effects are observed for the $p_{T,t}$ distribution, where the 
EW corrections tend to soften the tail, by up to $-25\%$ for $p_{T,t} \sim 3$~TeV. 
These effects are larger than the pure-QCD scale uncertainties and therefore will be 
important to include for the shape-modelling of this distribution, as the LHC experiments 
begin to measure the high tail with better statistical errors. 

It is well-known that the differential $t\bar{t}$ cross section contains large logarithms 
in regions of soft-gluon phase space and that in order to maintain perturbatively 
well-behaved predictions it is often necessary to resum these logarithms. 
This has been the subject of more than two decades of work, however, 
as reviewed in a separate talk \cite{ferroglia-top}, there has been recent progress on 
this front regarding the numerical implementation of soft-gluon resummation for boosted 
top-quark production. 
This is particularly relevant for the $M_{t\bar{t}}$ distribution where large logs 
$\sim \log(m_t/M_{t\bar{t}})$ are present in the tail. 
In the tail region, the resummation reduces the dependence on the choice of dynamical scale. 
There is ongoing work to consistently match the NNLO fixed order 
results with these soft-gluon resummed cross sections. 

NNLO results for distributions, including \texttt{fastNLO} tables and EW-rescaling 
factors are collected 
at \href{www.precision.hep.phy.cam.ac.uk/results}{http://www.precision.hep.phy.cam.ac.uk/results/}.

\vspace{-0.3cm}
\subsection{Application: constraining PDFs with distributions}

One of the lasting impacts of the top-quark physics program at the LHC will 
be the precise ways in which it can probe QCD. Perhaps the most important
of these will be providing non-trivial constraints on PDFs. 
Extensive investigations of the constraints provided by the inclusive $t\bar{t}$ cross section 
can be found for example, in ref.s~\cite{Czakon:2013tha,Alekhin:2017kpj}. 
A study exploring the extent to which the ATLAS and CMS 8 TeV differential
$t\bar{t}$ measurements, together with differential NNLO predictions,
can provide constraints beyond those provided by $t\bar{t}$ cross section measurements, 
was presented in ref.~\cite{Czakon:2016olj}.
The picture is promising, particularly for the gluon PDF at high-$x$, 
where including the differential top data tends to reduce the PDF-uncertainty bands
by better than a factor of 2. 
The study indicates that in this region, top data can be competitive with inclusive-jet
data, in terms of the potential constraints they can provide.

Another outcome of this study was the recommendation of the $y_t$ and $y_{t\bar{t}}$ 
normalized distributions from ATLAS and CMS respectively, as the differential
8 TeV measurements that are most suitable for use in a PDF fit. 
The reasons behind this seemingly eclectic choice were that for this combination 
of observables one can achieve good agreement between theoretical predictions and 
the measurements by both experiments, as well as good constraints on the PDFs. 
These distributions are also less sensitive to `typical' beyond-the-SM effects, which 
may show up as enhancements in the tails of $p_{T,t}$ and $M_{t\bar{t}}$, as well 
as to the uncertainty on the value of $m_t$.
Unfortunately, it was not possible to use multiple distributions from the same experiment
in the fitting, since experimental and theoretical systematics correlating distributions 
are not available. 
Once these correlations are known, this is certainly one aspect where there 
is potential to achieve further constraining power from measurements. 

\vspace{-0.4cm}
\section{Precision for top-pair production and decay} \label{sec:decayed}

\vspace{-0.2cm}
\subsection{Why bother with the decay?}

The presence of top quarks is always inferred via measurements of their 
decay products in phase space that is constrained by experimental detector 
geometries. 
In order to be compared against theoretical predictions for stable top quarks, 
this means that experimental measurements must (a) be extrapolated from the 
fiducial regions to the full phase space, and (b) be modelled from the 
particle-level back to some definition of `top partons.'
Because these extrapolations are performed using Parton Shower Monte Carlos (MCs), 
which include the top decay formally only at LO, an estimate of the systematic 
uncertainty introduced on the shape and normalization of such measurements, 
due to missing higher order corrections in the decay, is currently not possible. 
Since different MCs may have different parton shower algorithms and different 
methods of attaching the decay to the production subprocess, the very definition 
of a `top parton,' in the way this modelling is currently done, seems likely to 
be MC-dependent. 
To fully exploit the recent advances in precision at the level of stable
top quarks, these observations must be better understood, and the underlying 
systematics better controlled.    

Measurements presented at the level of the top-quark decay products can be 
compared without such extrapolations (and hence without this missing systematic
uncertainty) to theoretical predictions. 
In this section, recent work on including higher-order corrections to the 
decay subprocess as well as on including finite-width and non-resonant effects is
summarized. 
What emerges is a consistent picture; that higher-order corrections to the decay 
(particularly NLO) are generically important to describe regions of phase space
constrained by experimental selection cuts. 

\vspace{-0.2cm}
\subsection{High precision production and decay at fixed order}

The decay of the top quark is included in theoretical predictions either 
by treating the decaying top as strictly onshell -- the narrow-width approximation (NWA) -- 
or in offshell approaches, typically by using the complex-mass renormalization
scheme. 

\begin{figure}[htb]
\centering
\hspace{-0.2cm} \includegraphics[trim=0.3cm 0.0cm 0.25cm 0.7cm,clip,width=7.6cm]{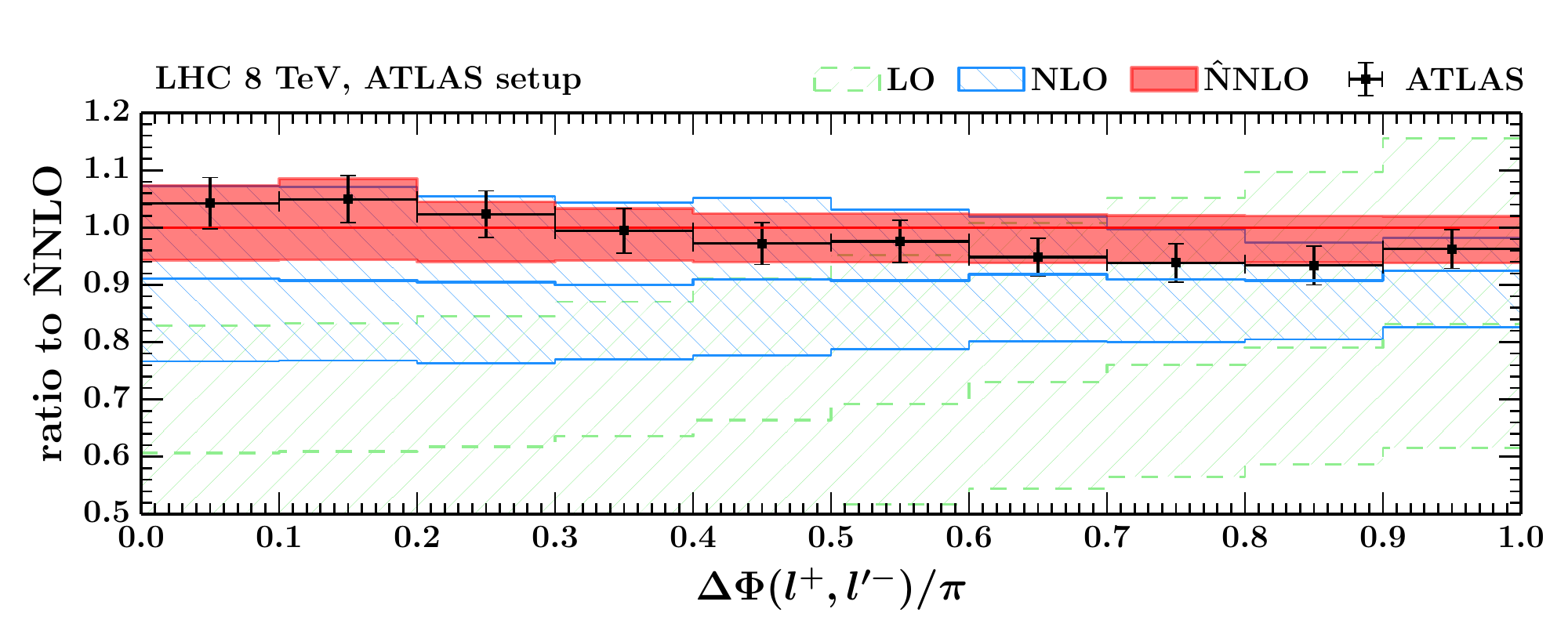} \hspace{-0.2cm} 
\includegraphics[trim=0.3cm 0.0cm 0.25cm 0.7cm,clip,width=7.6cm]{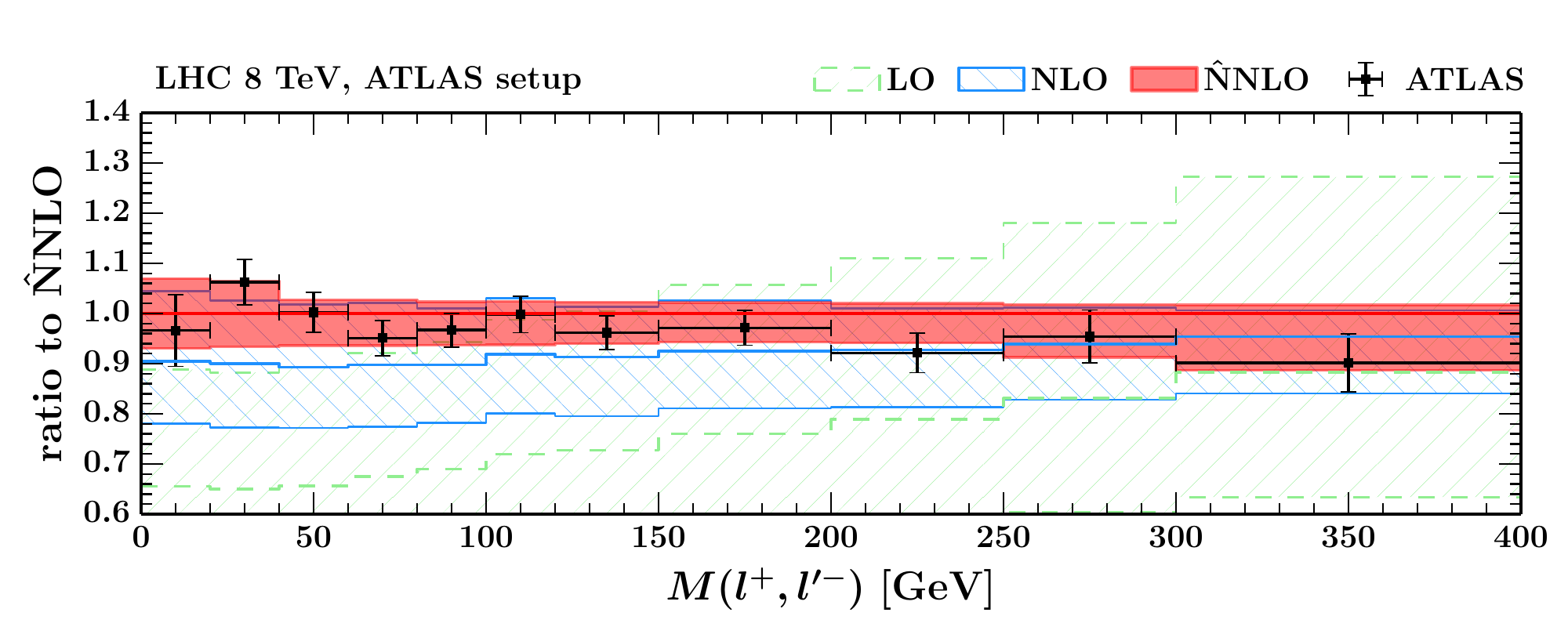} \\
\includegraphics[trim=0.3cm 0.0cm 0.25cm 1.3cm,clip,width=7.6cm]{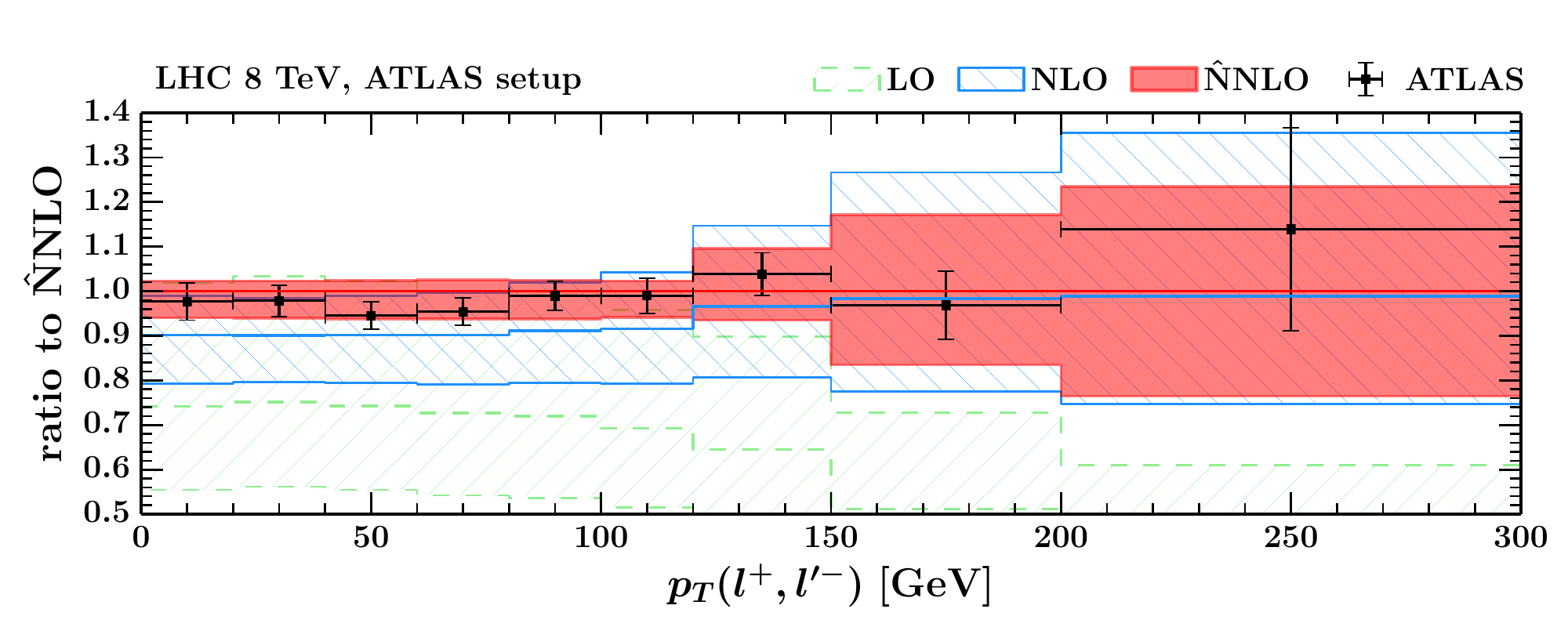}
\vspace{-0.3cm}
\caption{
Distributions of the azimuthal angle between the charge leptons, $\Delta\Phi(l^+,l'^-)$ and the
invariant mass, $M(l^+,l'^-)$, and transverse momentum, $p_T(l^+,l'^-)$, of the lepton-pair.
The plots show the ATLAS measurements \cite{Aaboud:2017ujq} as well as the LO, NLO and \^NNLO predictions normalized to \^NNLO.
The errorbars and shaded bands indicate the experimental and theoretical uncertainties respectively.
Comparisons to similar measurements by CMS \cite{Khachatryan:2015oqa} have been made in \cite{Gao:2017goi}.}
\label{fig:atlas-fid-ratios}
\end{figure}

In the NWA top-pair production was until recently known up to NLO \cite{Bernreuther:2004jv,Melnikov:2009dn,Campbell:2012uf}.
In these works the importance of including NLO corrections in the top-quark decay, 
was already pointed out. 
In recent work \cite{Gao:2017goi}, fully-differential predictions containing an approximation
to NNLO in production (with decays included up to NLO) \cite{Broggio:2014yca} were combined
with the differential decay at NNLO \cite{Gao:2012ja}, providing the most precise 
predictions at fixed-order, to date, for the di-lepton channel at 8 TeV. 
The best predictions are dubbed \^NNLO to distinguish the fact that the production 
includes an approximation to the exact NNLO. 
In fig.~\ref{fig:atlas-fid-ratios} these new predictions are compared to recently 
published ATLAS differential measurements \cite{Aaboud:2017ujq} in a fiducial region defined 
through selection cuts on charged leptons.  
There are two main messages to take from fig.~\ref{fig:atlas-fid-ratios}, namely that the 
corrections beyond NLO are significant $\sim 10\%$ and result in smaller scale variation 
uncertainties, and also that including them improves the comparison of theory to measurement. 

Ref.~\cite{Gao:2017goi} additionally highlighted that the size of higher-order corrections 
in the decay can depend on selection cuts. 
In particular, the size of these corrections and in particular the NLO, can be significant,
$\mathcal{O}(-10\%)$, when cuts are placed on $b$jets. 
This indicates that including the decay at LO alone is not sufficient to generically 
describe fiducial regions well. 
We point out that this is not an academic observation -- effects such as these from the decay 
will propagate through to measurements of the inclusive $t\bar{t}$ cross section, since 
the latter relies on extrapolations from measurements in fiducial regions.

While the NWA is an excellent approximation for a large class of observables, by construction
it does not capture finite-top-width effects and the effects of non-resonant contributions. 
As such, certain regions of phase-space, crucial for example the $W$-$b$jet invariant-mass 
and the edge of the lepton-$b$jet invariant mass, $M_{lb}$, are poorly described by the NWA. 
Furthermore, these effects are not only important to include for measurements of top-pair 
production, but additionally play a role in accurately describing top-quark backgrounds
to processes such as $H\to W W^{(*)}$ \cite{Frederix:2013gra,Cascioli:2013wga}.
The past years have seen a serious effort to include these effects, 
up to NLO for the process of $t\bar{t}$ in the dilepton channel  
\cite{Bevilacqua:2010qb,Denner:2010jp,Denner:2012yc,Falgari:2013gwa,Heinrich:2013qaa,Frederix:2013gra,Cascioli:2013wga}.
Very recently the hugely complex fully-offshell calculations of $t\bar{t}$ in the 
lepton+jets channel \cite{Denner:2017kzu} and of 
$t\bar{t}j$ \cite{Bevilacqua:2015qha,Bevilacqua:2016jfk} at NLO-QCD have been completed. 

An in-depth discussion of the findings of all these works is beyond the scope of these
proceedings. What we emphasise here is that, in support of the discussion above, 
where the NWA \emph{is} a good approximation, this is only generally the case when 
corrections to the decay subprocesses are also included. 
This is not only found for $t\bar{t}$, but also in the recent NWA vs offshell comparison
in ref.~\cite{Bevilacqua:2017ipv} for $t\bar{t}j$ where again it was shown that
including NLO production, but not decay, corrections has a detrimental effect to the 
quality of description the NWA provides. 

Finally, NLO-EW corrections are now known for offshell $t\bar{t}$ production in the di-lepton 
channel \cite{Denner:2016jyo}.
As expected, at the inclusive level, EW corrections are found to be small, below a per-cent. 
However, as with onshell $t\bar{t}$, significant negative EW corrections are observed in the 
high-$p_T$ tails of the leptons or $b$jets, reaching $-15\%$ for values of $p_T \simeq 800$~GeV. 

\vspace{-0.2cm}
\subsection{NLO+PS for offshell $t\bar{t}$}

Translating the progress in including the decay at fixed order, to predictions 
matched to parton showers has not been straightforward.  
This required the development of a modified NLO+PS matching scheme \cite{Jezo:2015aia}, 
and for $t\bar{t}$ this was subsequently applied to the offshell process in the 
di-lepton channel \cite{Jezo:2016ujg}.
This work and ongoing developments are discussed in more detail in these proceedings \cite{jezo-top}.
In the spirit of the rest of this section, we would like to highlight one aspect, 
namely the comparisons between 
available MCs with different underlying approximations for the hard matrix elements. 
Unsurprisingly, the MC with underlying NLO production-and-decay NWA matrix 
elements approximates the fully-offshell MC better than the commonly-used MC with 
an underlying $t\bar{t}$ with NLO corrections in production. 
This further cements the pattern that decay corrections are a crucial ingredient
to accurately model top-quark final states in fiducial regions. 
We also point to ongoing work within the \texttt{HW7} collaboration on including 
full NLO top-decays in their shower framework \cite{Bellm:2017idv}. 
It is important to understand whether, compared to using MCs which use stable $t\bar{t}$ matrix 
elements, this improved perturbative modelling has an impact on top-quark mass extractions. 
This was discussed in a dedicated talk at this conference \cite{nason-top}.

As a final remark, as discussed in the sec.~\ref{sec:stable}, the theoretical 
description of stable $t\bar{t}$ production is at a very high level of precision and sophistication. 
This has the potential for hugely impactful applications, such as PDF-extractions. 
However, to fully exploit these theoretical predictions it is vital to understand the 
dynamics of top-quark final states and systematically study the systematics associated 
with extrapolating from particle to parton level. 
Sec.~\ref{sec:decayed} highlighted that the tools to do this at high precision are 
already becoming available, and particular importance is placed on the inclusion of 
corrections to the decay as well as to the production subprocesses.

\vspace{-0.4cm}
\Acknowledgements
I would like to extend a big thank you to the organising committee for the invitation 
to give this talk at this wonderful conference.
I would also like to thank A.~Broggio, M.~Czakon, J.~Gao, D.~Heymes, A.~Mitov, and A.~Signer
for collaboration and many insightful and useful discussions.


\begin{thebibliography}{99}

\bibitem{Czakon:2015owf}
M.~Czakon, D.~Heymes and A.~Mitov, 
  \href{http://dx.doi.org/10.1103/PhysRevLett.116.082003}{\emph{Phys. Rev.
  Lett.} {\bf 116} (2016) 082003}, [\href{http://arxiv.org/abs/1511.00549}{{\tt
  1511.00549}}].

\bibitem{Czakon:2016ckf}
M.~Czakon, P.~Fiedler, D.~Heymes and A.~Mitov, 
  \href{http://dx.doi.org/10.1007/JHEP05(2016)034}{\emph{JHEP} {\bf 05} (2016)
  034}, [\href{http://arxiv.org/abs/1601.05375}{{\tt 1601.05375}}].

\bibitem{Czakon:2016dgf}
M.~Czakon, D.~Heymes and A.~Mitov, 
  \href{http://dx.doi.org/10.1007/JHEP04(2017)071}{\emph{JHEP} {\bf 04} (2017)
  071}, [\href{http://arxiv.org/abs/1606.03350}{{\tt 1606.03350}}].

\bibitem{Czakon:2014oma}
M.~Czakon and D.~Heymes, 
  \href{http://dx.doi.org/10.1016/j.nuclphysb.2014.11.006}{\emph{Nucl. Phys.}
  {\bf B890} (2014) 152--227}, [\href{http://arxiv.org/abs/1408.2500}{{\tt
  1408.2500}}].

\bibitem{Britzger:2012bs}
{\scshape fastNLO} collaboration, D.~Britzger, K.~Rabbertz, F.~Stober and
  M.~Wobisch, 
  \emph{{Proceedings, 20th International Workshop on Deep-Inelastic Scattering
  and Related Subjects (DIS 2012): Bonn, Germany, 2012}}
\newblock \href{http://arxiv.org/abs/1208.3641}{{\tt 1208.3641}}.

\bibitem{Czakon:2017dip}
M.~Czakon, D.~Heymes and A.~Mitov, 
  \href{http://arxiv.org/abs/1704.08551}{{\tt 1704.08551}}.

\bibitem{Czakon:2017wor}
M.~Czakon, et~al.,
  \href{http://dx.doi.org/10.1007/JHEP10(2017)186}{\emph{JHEP} {\bf 10} (2017)
  186}, [\href{http://arxiv.org/abs/1705.04105}{{\tt 1705.04105}}].

\bibitem{Czakon:2017lgo}
M.~Czakon, et~al.,
  \href{http://arxiv.org/abs/1711.03945}{{\tt 1711.03945}}.

\bibitem{pagani-top}
M.~Czakon, et~al.,
  \emph{{These proceedings}}.
\newblock \href{http://arxiv.org/abs/1712.04842}{{\tt 1712.04842}}.

\bibitem{ferroglia-top}
A.~Ferroglia, \emph{{These proceedings}}.

\bibitem{Czakon:2013tha} 
  M.~Czakon, M.~L.~Mangano, A.~Mitov and J.~Rojo,
  \href{http://dx.doi.org/10.1007/JHEP07(2013)167}{\emph{JHEP} {\bf 1307}, 167 (2013)},
  [\href{http://arxiv.org/abs/1303.7215}{\tt 1303.7215}].

\bibitem{Alekhin:2017kpj} 
  S.~Alekhin, J.~Blümlein, S.~Moch and R.~Placakyte,
  \href{http://dx.doi.org/10.1103/PhysRevD.96.014011}{\emph{Phys.\ Rev.\ D} {\bf 96}, no. 1, 014011 (2017)},
  [\href{http://arxiv.org/abs/1701.05838}{\tt 1701.05838}].

\bibitem{Czakon:2016olj}
M.~Czakon, N.~P. Hartland, A.~Mitov, E.~R. Nocera and J.~Rojo, 
  \href{http://dx.doi.org/10.1007/JHEP04(2017)044}{\emph{JHEP} {\bf 04} (2017)
  044}, [\href{http://arxiv.org/abs/1611.08609}{{\tt 1611.08609}}].

\bibitem{Bernreuther:2004jv}
W.~Bernreuther, A.~Brandenburg, Z.~G. Si and P.~Uwer, 
  \href{http://dx.doi.org/10.1016/j.nuclphysb.2004.04.019}{\emph{Nucl. Phys.}
  {\bf B690} (2004) 81--137}, [\href{http://arxiv.org/abs/hep-ph/0403035}{{\tt
  hep-ph/0403035}}].

\bibitem{Melnikov:2009dn}
K.~Melnikov and M.~Schulze, 
  \href{http://dx.doi.org/10.1088/1126-6708/2009/08/049}{\emph{JHEP} {\bf 08}
  (2009) 049}, [\href{http://arxiv.org/abs/0907.3090}{{\tt 0907.3090}}].

\bibitem{Campbell:2012uf}
J.~M. Campbell and R.~K. Ellis, 
  \href{http://dx.doi.org/10.1088/0954-3899/42/1/015005}{\emph{J.
  Phys.} {\bf G42} (2015) 015005}, [\href{http://arxiv.org/abs/1204.1513}{{\tt
  1204.1513}}].

\bibitem{Gao:2017goi}
J.~Gao and A.~S. Papanastasiou, 
  \href{http://dx.doi.org/10.1103/PhysRevD.96.051501}{\emph{Phys. Rev.} {\bf
  D96} (2017) 051501}, [\href{http://arxiv.org/abs/1705.08903}{{\tt
  1705.08903}}].

\bibitem{Broggio:2014yca}
A.~Broggio, A.~S. Papanastasiou and A.~Signer, 
  \href{http://dx.doi.org/10.1007/JHEP10(2014)098}{\emph{JHEP} {\bf 10} (2014)
  98}, [\href{http://arxiv.org/abs/1407.2532}{{\tt 1407.2532}}].

\bibitem{Gao:2012ja}
J.~Gao, C.~S. Li and H.~X. Zhu, 
  \href{http://dx.doi.org/10.1103/PhysRevLett.110.042001}{\emph{Phys. Rev.
  Lett.} {\bf 110} (2013) 042001}, [\href{http://arxiv.org/abs/1210.2808}{{\tt
  1210.2808}}].

\bibitem{Aaboud:2017ujq}
{\scshape ATLAS} collaboration, M.~Aaboud et~al., 
  \href{http://dx.doi.org/10.1140/epjc/s10052-017-5349-9}{\emph{Eur. Phys. J.}
  {\bf C77} (2017) 804}, [\href{http://arxiv.org/abs/1709.09407}{{\tt
  1709.09407}}].

\bibitem{Khachatryan:2015oqa}
{\scshape CMS} collaboration, V.~Khachatryan et~al., 
  \href{http://dx.doi.org/10.1140/epjc/s10052-015-3709-x}{\emph{Eur. Phys. J.}
  {\bf C75} (2015) 542}, [\href{http://arxiv.org/abs/1505.04480}{{\tt
  1505.04480}}].

\bibitem{Frederix:2013gra}
R.~Frederix, 
  \href{http://dx.doi.org/10.1103/PhysRevLett.112.082002}{\emph{Phys. Rev.
  Lett.} {\bf 112} (2014) 082002}, [\href{http://arxiv.org/abs/1311.4893}{{\tt
  1311.4893}}].

\bibitem{Cascioli:2013wga}
F.~Cascioli, S.~Kallweit, P.~Maierh{\" o}fer and S.~Pozzorini, 
  \href{http://dx.doi.org/10.1140/epjc/s10052-014-2783-9}{\emph{Eur. Phys. J.}
  {\bf C74} (2014) 2783}, [\href{http://arxiv.org/abs/1312.0546}{{\tt
  1312.0546}}].

\bibitem{Bevilacqua:2010qb}
G.~Bevilacqua, M.~Czakon, A.~van Hameren, C.~G. Papadopoulos and M.~Worek,
  \href{http://dx.doi.org/10.1007/JHEP02(2011)083}{\emph{JHEP} {\bf 02} (2011)
  083}, [\href{http://arxiv.org/abs/1012.4230}{{\tt 1012.4230}}].

\bibitem{Denner:2010jp}
A.~Denner, S.~Dittmaier, S.~Kallweit and S.~Pozzorini, 
  \href{http://dx.doi.org/10.1103/PhysRevLett.106.052001}{\emph{Phys. Rev.
  Lett.} {\bf 106} (2011) 052001}, [\href{http://arxiv.org/abs/1012.3975}{{\tt
  1012.3975}}].

\bibitem{Denner:2012yc}
A.~Denner, S.~Dittmaier, S.~Kallweit and S.~Pozzorini, 
  \href{http://dx.doi.org/10.1007/JHEP10(2012)110}{\emph{JHEP} {\bf 10} (2012)
  110}, [\href{http://arxiv.org/abs/1207.5018}{{\tt 1207.5018}}].

\bibitem{Falgari:2013gwa}
P.~Falgari, A.~S. Papanastasiou and A.~Signer, 
  \href{http://dx.doi.org/10.1007/JHEP05(2013)156}{\emph{JHEP} {\bf 05} (2013)
  156}, [\href{http://arxiv.org/abs/1303.5299}{{\tt 1303.5299}}].

\bibitem{Heinrich:2013qaa}
G.~Heinrich, A.~Maier, R.~Nisius, J.~Schlenk and J.~Winter, 
  \href{http://dx.doi.org/10.1007/JHEP06(2014)158}{\emph{JHEP} {\bf 06} (2014)
  158}, [\href{http://arxiv.org/abs/1312.6659}{{\tt 1312.6659}}].

\bibitem{Denner:2017kzu}
A.~Denner and M.~Pellen, 
  \href{http://arxiv.org/abs/1711.10359}{{\tt 1711.10359}}.

\bibitem{Bevilacqua:2015qha}
G.~Bevilacqua, H.~B. Hartanto, M.~Kraus and M.~Worek, 
  \href{http://dx.doi.org/10.1103/PhysRevLett.116.052003}{\emph{Phys. Rev.
  Lett.} {\bf 116} (2016) 052003}, [\href{http://arxiv.org/abs/1509.09242}{{\tt
  1509.09242}}].

\bibitem{Bevilacqua:2016jfk}
G.~Bevilacqua, H.~B. Hartanto, M.~Kraus and M.~Worek, 
  \href{http://dx.doi.org/10.1007/JHEP11(2016)098}{\emph{JHEP} {\bf 11} (2016)
  098}, [\href{http://arxiv.org/abs/1609.01659}{{\tt 1609.01659}}].

\bibitem{Bevilacqua:2017ipv}
G.~Bevilacqua, H.~B. Hartanto, M.~Kraus, M.~Schulze and M.~Worek, 
  \href{http://arxiv.org/abs/1710.07515}{{\tt 1710.07515}}.

\bibitem{Denner:2016jyo}
A.~Denner and M.~Pellen, 
  \href{http://dx.doi.org/10.1007/JHEP08(2016)155}{\emph{JHEP} {\bf 08} (2016)
  155}, [\href{http://arxiv.org/abs/1607.05571}{{\tt 1607.05571}}].

\bibitem{Jezo:2015aia}
T.~Je{\v z}o and P.~Nason, 
  \href{http://dx.doi.org/10.1007/JHEP12(2015)065}{\emph{JHEP} {\bf 12} (2015)
  065}, [\href{http://arxiv.org/abs/1509.09071}{{\tt 1509.09071}}].

\bibitem{Jezo:2016ujg}
T.~Je{\v z}o, J.~M. Lindert, P.~Nason, C.~Oleari and S.~Pozzorini, 
  \href{http://dx.doi.org/10.1140/epjc/s10052-016-4538-2}{\emph{Eur. Phys. J.}
  {\bf C76} (2016) 691}, [\href{http://arxiv.org/abs/1607.04538}{{\tt
  1607.04538}}].

\bibitem{Bellm:2017idv}
J.~Bellm, et~al.,
  \emph{{These proceedings}}.
  \href{http://arxiv.org/abs/1711.11570}{{\tt 1711.11570}}.

\bibitem{jezo-top}
T.~Je{\v z}o, \emph{{These proceedings}}.

\bibitem{nason-top}
P.~Nason, \emph{{These proceedings}}.

\end{thebibliography}
\end{document}